\documentclass[journal]{IEEEtran}
\usepackage{url}
\usepackage{hyperref}
\usepackage{graphicx}
\usepackage{subfigure}
\usepackage{subfigure}
\usepackage{multirow}
\usepackage{color, colortbl}

\definecolor{mydarkgreen}{RGB}{0,160,0}

\usepackage{colortbl}
\usepackage[table,xcdraw]{xcolor}
\usepackage{cite}
\usepackage{soul}

\usepackage{bbding}
\usepackage{tikz}

\begin{document}
%
\title{Characterizing Continual Learning Scenarios and Strategies for Audio Analysis}
%



\author{Ruchi Bhatt, Pratibha Kumari, Dwarikanath Mahapatra, Abdulmotaleb El Saddik and Mukesh Saini

\thanks{The authors Ruchi Bhatt is with the Department of Computer Science \& Engineering, Indian Institute of Technology Ropar, Punjab, India (email: ruchi.21csz0007@iitrpr.ac.in), Pratibha Kumari is with the Faculty of Informatics and Data Science, University of Regensburg, Regensburg, Germany (email: Pratibha.Kumari@ur.de), Dwarikanath Mahapatra is with Inception Institute of AI, Abu Dhabi, United Arab Emirates (email: dwarikanath.mahapatra@inceptioniai.org), Abdulmotaleb El Saddik is with Mohamed Bin Zayed University of Artificial Intelligence (MBZUAI), Abu Dhabi, The United Arab Emirates \& School of Electrical Engineering and Computer Science University of Ottawa, Canada (email: elsaddik@uottawa.ca), and Mukesh Saini is with the Department of Computer Science \& Engineering, Indian Institute of Technology Ropar, Punjab, India \& Mohamed Bin Zayed University of Artificial Intelligence (MBZUAI), Abu Dhabi, The United Arab Emirates (email:mukesh@iitrpr.ac.in)}
}



\maketitle
\begin{abstract}
Audio analysis is useful in many application scenarios. The state-of-the-art audio analysis approaches assume the data distribution at training and deployment time will be the same. However, due to various real-life challenges, the data may encounter drift in its distribution or can encounter new classes in the late future. Thus, a one-time trained model might not perform adequately. Continual learning (CL) approaches are devised to handle such changes in data distribution. There have been a few attempts to use CL approaches for audio analysis. Yet, there is a lack of a systematic evaluation framework. In this paper, we create a comprehensive CL dataset and characterize CL approaches for audio-based monitoring tasks. We have investigated the following CL and non-CL approaches: EWC, LwF, SI, GEM, A-GEM, GDumb, Replay, Naive, Cumulative, and Joint training. The study is very beneficial for researchers and practitioners working in the area of audio analysis for developing adaptive models.
We observed that Replay achieved better results than other methods in the DCASE challenge data. It achieved an accuracy of 70.12\% for the domain incremental scenario and an accuracy of 96.98\% for the class incremental scenario.


\end{abstract}


%
\IEEEpeerreviewmaketitle


\section{Introduction}

Audio plays an important role in machine learning and deep learning applications, contributing to various tasks such as event detection~\cite{choudhary2022low}, speech recognition \cite{nassif2019speech}, audio surveillance \cite{crocco2016audio}, sound classification \cite{l2_wang2019continual}, audio signal processing \cite{purwins2019deep}, music generation, and environmental sound analysis. In this paper, we characterize continual learning (CL) approaches for audio analysis with a case study of audio-based monitoring.  Audio-based monitoring offers distinct advantages such as low-cost deployment, non-invasiveness, remote operation, and the ability to capture rich contextual information~\cite{choudhary2022fingerprinting,choudhary2022audio,bhatt2022experimental}. With advancements in sensor technologies and the widespread availability of audio data, audio-based monitoring has become increasingly prevalent across various domains, including public surveillance, environmental sound monitoring, wildlife conservation, and healthcare. 

State-of-the-art audio analysis models follow one-time training with all the available data and employ the trained model for all future inferences. The assumption of prior availability of all the data to train a deep model is not practical. Data usually comes over time in real-world environments, and the model needs to be updated with the new batch of data~\cite{kumari2022anomaly,kumari2024concept}. However, as the deep models learn new concepts from newly available data, they tend to forget the past learning and hence perform poorly on past data, leading to a phenomenon known as catastrophic forgetting \mbox{\cite{shao2022overcoming,french1999catastrophic,mccloskey1989catastrophic,ratcliff1990connectionist}}. To mitigate the effect of catastrophic forgetting, a human-like lifelong learning paradigm, popularly termed as Continual Learning (CL) is gaining major attention recently. CL offers to acquire new skills and accumulate knowledge over time without the risk of forgetting previously learned experiences. Consequently, several CL strategies have been proposed in various research fields, including object detection~\mbox{\cite{menezes2023continual}}, robotics~\mbox{\cite{lesort2020continual}}, medical image analysis~\mbox{\cite{derakhshani2022lifelonger,kumari2023continual}}, etc. 

Especially in the audio domain, CL becomes pertinent due to the streaming and dynamically evolving nature of sound environments. By continuously updating the deployed model with new audio samples, a CL strategy can be leveraged to recognize and classify new audio events, adapt to new acoustic conditions, and generalize across different audio contexts. 
CL has been exploited in audio applications including sound classification \cite{l2_wang2019continual}, audio analysis/classification \cite{l4_wang2021few,l10_karam2022task,muthuchamy2023adapter,zhang2023remember,l17_mulimani2024class}, sound event detection \cite{wang2020few,l3_koh2020incremental}, audio captioning \cite{l6_berg2021continual}, audio-visual learning \cite{Mo_2023_ICCV,l11_mo2023class,l12_pian2023audio}, fake audio detection \cite{l5_ma2021continual}, speech recognition \cite{sadhu2020continual}, etc. 
Nevertheless, efforts to adopt CL strategies in the audio monitoring application are still in their early stage. The authors mostly use custom datasets in ad-hoc scenarios to demonstrate their performance. There is a need for a systematic evaluation framework and scenario-rich dataset to assess the current state of CL works in the field of audio analysis. 

We curate a sequential dataset in ``domain-shift'' as well as ``new-class'' situations with different CL scenarios using data from DCASE challenges\footnote{https://dcase.community/}. DCASE is a huge publicly available audio dataset repository with annotations. Although Zhou et al.~\cite{l18_zhou2024anomaly} utilize anomaly dataset from the DCASE2020 task2 challenge, they consider drift as a change of machine type in the anomaly detection application, whereas we considered naturally occurring drifts such as weather conditions, machine load over time, etc. Further, we identify and benchmark popular CL strategies for audio-based monitoring application including EWC \cite{kirkpatrick2017overcoming}, SI \cite{zenke2017continual}, LwF \cite{MALTONI201956}\cite{li2017learning}, GEM \cite{lopez2017gradient}, A-GEM \cite{chaudhry2018efficient}, GDumb \cite{prabhu2020gdumb}, Replay\cite{rolnick2019experience}, Naive, Cumulative, and Joint training. We conduct a thorough analysis of the performance and examine how well catastrophic forgetting is controlled by these CL strategies in different dynamically changing audio environment settings.


The major contributions of this paper are as follows: 
\begin{itemize}
\item We have given a comparative study of different CL approaches, namely EWC, LwF, SI, GEM, A-GEM, GDumb, Replay, Naive, Cumulative, and Joint training for audio-based monitoring tasks.


\item We have prepared a domain and class incremental scenarios dataset using the publicly available DCASE challenge datasets from the years 2020 to 2023. 
\end {itemize}

The rest of the paper is structured as follows. Section \ref{sec:CLbasics} gives the details of CL methods and scenarios. Section \ref{sec:baslines} consists of the basics of CL and Non-CL baselines. Section \ref{sec:dataset} includes a detailed description of the dataset used. Section \ref{sec:experiments} consists of the experimental setup and results, followed by a discussion in section \ref{sec:dicsussion}, and the conclusion in section \ref{sec:conclusion}.

\begin{table*}[!ht]
\centering
\caption{Summary of state-of-the-art continual learning works in audio domain}
\label{tab:comp_literature}

\begin{tabular}{|c|c|c|c|c|}
\hline
\multicolumn{1}{|c|}{\textbf{Work}} 
& \multicolumn{1}{c|}{\textbf{CL Strategy}} 
& \multicolumn{1}{c|}{\textbf{CL Scenario}}  
& \multicolumn{1}{c|}{\textbf{Dataset}} &
\textbf{\begin{tabular}[c]{@{}c@{}}Application (shift detail if any)\end{tabular}}
 \\ \hline

 \rowcolor{gray!10}
Wang et. al.\cite{l2_wang2019continual} & \begin{tabular}[c]{@{}l@{}}Rehearsal \end{tabular} & CI  & ESC10 & Audio scene classification \\ 
 \rowcolor{gray!10}
Koh et. al.\cite{l3_koh2020incremental} & Architectural & CI  & DCASE16, US-SED & Audio scene classification \\ 
 \rowcolor{gray!10}
Kwon et. al.\cite{l8_kwon2021fasticarl} & Rehearsal & CI  & \begin{tabular}[c]{@{}l@{}}EmotionSense,\\ UrbanSound8K\end{tabular} & Audio scene classification \\ 
 \rowcolor{gray!10}
Wang et. al.\cite{l9_wang2022learning} & Regularization & CI  & \begin{tabular}[c]{@{}l@{}}UrbanSound8K, VGGSound,\\ urban acoustic scenes 2019 \\ (DCASE TAU19)\end{tabular} & Audio scene classification \\ 
 \rowcolor{gray!10}
Karam et. al.\cite{l10_karam2022task} & Rehearsal & TI  & ESC50, UrbanSound8K & Audio scene classification \\ 
 \rowcolor{gray!10}
Mulimani et. al.\cite{l14_mulimani2023incremental} & Regularization & CI  & \begin{tabular}[c]{@{}l@{}}DCASE TUT acoustic scenes\\ 2017, TUT 2016/2017\end{tabular} & Audio scene classification \\ 
 \rowcolor{gray!10}
Sun et. al.\cite{l16_sun2023environmental} & \begin{tabular}[c]{@{}l@{}}Rehearsal \end{tabular} & CI  & ESC10, ESC50, UrbanSound8K & Audio scene classification \\ \hline

Wang et. al.\cite{l4_wang2021few} & Architectural & CI  & AudioSet, ESC50 & \begin{tabular}[c]{@{}l@{}}Multi-label audio classification\end{tabular} \\ 
Wang et. al.\cite{l7_wang2021calls} & Architectural & CI  & \begin{tabular}[c]{@{}l@{}}FSD-MIX-CLIPS (1s),  \\ FSD-MIX-SED (10s)\end{tabular} & Multi-label audio classification \\ 
 Mulimani et. al.\cite{l17_mulimani2024class} & Regularization & CI  & AudioSet & \begin{tabular}[c]{@{}l@{}}Multi-label audio classification\end{tabular} \\ \hline

 \rowcolor{gray!10}
Mo et. al.\cite{l11_mo2023class} & \begin{tabular}[c]{@{}l@{}}Regularization \\ \& Rehearsal\end{tabular} & CI  & \begin{tabular}[c]{@{}l@{}}VGGSound-Instruments,\\ VGGSound-100,\\ VGG-Sound Sources\end{tabular} & \begin{tabular}[c]{@{}l@{}}Multi-modal classification (audio,video)\end{tabular} \\ 
 \rowcolor{gray!10}
Pian et.al.\cite{l12_pian2023audio} & Rehearsal & CI  & \begin{tabular}[c]{@{}l@{}}AVE-CI, K-S-CI,\\ VS100-CI\end{tabular} & \begin{tabular}[c]{@{}l@{}}Multi-modal classification (audio,video)\end{tabular} \\ 
 \rowcolor{gray!10}
Kim et. al.\cite{l13_kim2024multi} & Regularization & TI  & \begin{tabular}[c]{@{}l@{}}AudioSet, VGGSound, MACS,\\ FSD50K, ClothoV2, AudioCaps\end{tabular} & \begin{tabular}[c]{@{}l@{}}Multi-modal classification (audio,video, text)\end{tabular} \\ \hline

Berg et. al.\cite{l6_berg2021continual} & Regularization & DI & Clotho, AudioCaps & \begin{tabular}[c]{@{}l@{}}Automated audio captioning \\(Audio clips are annotated with 5 captions\\ to each 10-30 sec each of both datasets) \end{tabular} \\ \hline

 \rowcolor{gray!10}
Ma et. al.\cite{l5_ma2021continual} & Regularization & DI & ASVSpoof2019 &  \begin{tabular}[c]{@{}l@{}}Fake audio detection\\ (Considered 2 subsets of the dataset \& \\selected 4 spoofing attacks with big\\ differences, from each subset separately)\end{tabular}  \\ \hline

Zhou et. al.\cite{l18_zhou2024anomaly} & Regularization & DI  & DCASE 2020 challenge Task2 & \begin{tabular}[c]{@{}l@{}}Anomaly sound detection \\(Considered subsequent machine type as drift)\end{tabular}  \\ \hline

\end{tabular}
\end{table*}

\section{Continual learning}\label{sec:CLbasics}

CL is a rapidly emerging research direction to mitigate catastrophic forgetting and mimic the human way of learning in deep neural networks \cite{de2021continual}. It refers to the learning strategy in which a machine continuously accumulates knowledge from the incoming data. Unlike traditional deep learning, where the complete data is accessible for training at the start, in CL, the data comes sequentially, also known as tasks or episodes. CL allows the acquisition of new knowledge while retaining the existing knowledge. 
However, there is a tradeoff between how well a model can retain past knowledge and learn new things. It is called the stability-plasticity dilemma \cite{mermillod2013stability}. The plasticity shows the ability of the model to learn new tasks, and the stability refers to maintaining the performance of previously learned tasks upon learning new concepts~\cite{Kim_2023_CVPR}. Efforts to increase either of them hinder the other, and vice versa \cite{wang2024comprehensive}. More favor for either plasticity or stability also depends on the targeted application. 

Various CL strategies have been formulated in literature to mitigate forgetting and facilitate knowledge accumulation. Fig.~{\ref{fig:flowchart} details broader modules of a CL framework, including CL scenario, CL data stream, CL strategy, evaluation scheme, and finally, available baselines. Depending upon what kind of change is expected, i.e., new class, new instances, shifted distribution, entirely new task, etc., there can be various settings, also known as CL scenarios. Once the possible change in data is identified for the application, a dataset stream containing multiple tasks, each having a train and test split, is prepared. Then, a suitable CL strategy from various categories is employed in the deep model to refrain from a drop in performance on past episodes upon learning new episodes. Lastly, the effectiveness of a particular CL strategy is analyzed against other CL and non-CL strategies with the help of specifically designed metrics to measure forgetting and knowledge transfer. We discuss major CL scenarios and strategies in sections {\ref{sec:CL_scenario}} and {\ref{sec:CL_technique}}, respectively.

\subsection{Continual learning scenarios}\label{sec:CL_scenario}
In continual learning, a ``scenario'' is the specific way in which new tasks are presented to the model over time. It describes the conditions under which the model learns and adapts, simulating how learning happens in the real world, where new information comes in a sequence. In analysis, particularly audio-based monitoring using anomaly detection, two situations are most prevalent: new samples of the existing class with various shifts and new intruder/event classes. 
Based on these situations, various CL scenarios are defined across multiple research domains~\cite{kumari2023continual}, out of which two majorly important and widely used scenarios are discussed below.

\subsubsection{Domain-incremental scenario}
Domain-incremental (DI) scenario deals with situations where the job remains the same, but the context or the information presented changes over time due to the non-stationary environment. 
The data distribution changes, but the classes remain the same for all subsequent tasks. The domain-incremental scenario \cite{mirza2022efficient} allows us to continuously update the model with new streams of data from shifted distributions. Examples of various shift sources in monitoring include variations due to weather, seasons, geo-location, surrounding noise levels across the day, etc.
\begin{table}[ht!]
\centering
\caption{Summary of major CL scenarios}
\label{tab:dis_cis}
\begin{tabular}{|l|l|}
\hline
\multicolumn{1}{|c|}{\textbf{CL Scenario}} & \multicolumn{1}{c|}{\textbf{Summary}} \\ \hline
DI & \begin{tabular}[c]{@{}l@{}}Each task contains new data distribution with the\\ same classes per task, task ID is not provided\end{tabular} \\ \hline
CI & \begin{tabular}[c]{@{}l@{}}Each task contains a new set of classes per task, \\task ID is inferred\end{tabular} \\ \hline
\end{tabular}
\end{table}


\subsubsection{Class-incremental scenario}
Class-incremental (CI) scenario \cite{tao2020few} is a scenario with a single job over the subsequent tasks. It deals with continuously extending the model to unseen classes with new streams of data. Usually, the classes in subsequent tasks are disjoint; hence, the knowledge to learn is also non-overlapping, causing this scenario to be the hardest among CL scenarios. An example of class-incremental learning in real-world monitoring is to identify a cat and dog in one episode, and then a cow and horse in the next episode \cite{van2022three}.

There are other CL scenarios available in the literature like Data-incremental, Task-incremental, and Hybrid scenarios. The data-incremental scenario allows us to continuously learn from new streams of data from an identical underlying distribution during sequential model training. It does not consider new classes or considerable shifts of data distribution and thus can also be considered as a domain-incremental scenario where the data distribution remains unchanged. The task-incremental (TI) scenario refers to learning different kinds of jobs over subsequent tasks. The jobs may also be highly uncorrelated. Each job is observed as a task, and it usually has a disjoint label space. The task-incremental scenario can be considered the class-incremental scenario where the subsequent tasks are observed as dissimilar jobs and classes associated with the tasks are disjoint. It can also be considered the domain-incremental scenario where the subsequent tasks are observed as similar jobs, and classes associated with them are unchanged.  A real-world example of task-incremental learning is the sequential learning of different sports or musical instruments over time. In the real world, it is highly unrealistic that the model will always encounter one type of incremental scenario. It is quite possible that the model can observe new classes as well as domain shifts. The model developed for handling class-incremental situations may not perform well for domain-incremental situations. Thus, there are attempts in the literature to evaluate the same model on a mix of incremental scenarios called hybrid scenarios~\cite{chen2019hybrid}. The two major CL scenarios are summarized in table \ref{tab:dis_cis}. 

\begin{figure*}[!ht]
\centering
\includegraphics[scale=0.8]{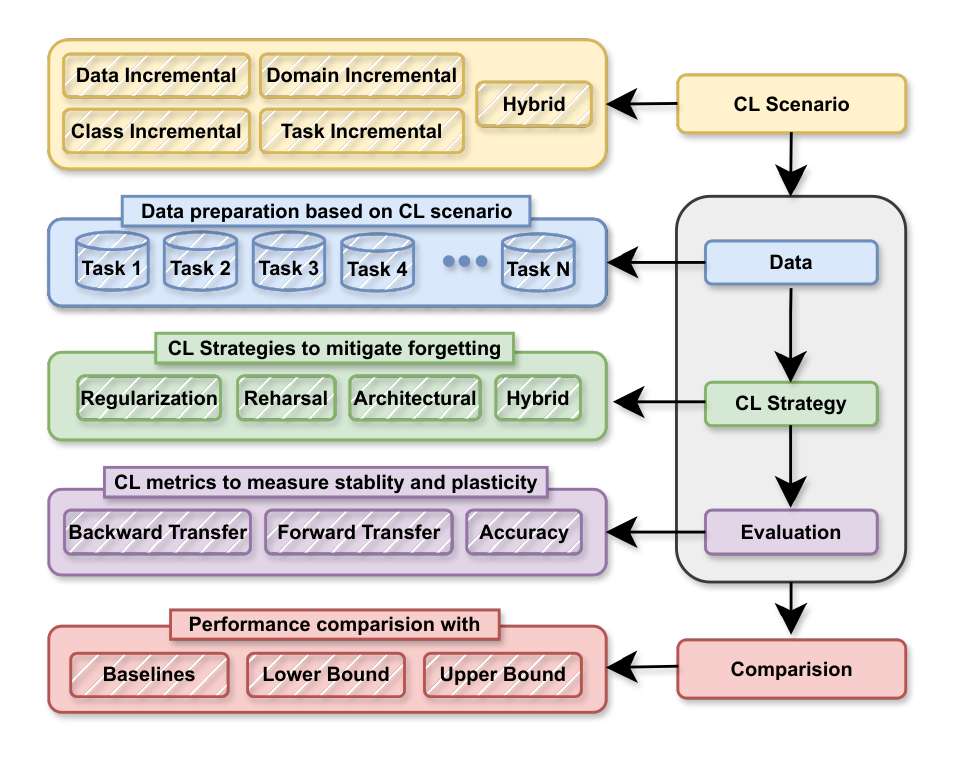}
\caption{Overview of a continual learning pipeline. This figure shows the available possible CL scenarios, the data preparation for different scenarios in subsets called tasks, broad categories of CL strategies based on literature, the well-known metrics used to evaluate the performances, and different ways of comparative analysis.}
\label{fig:flowchart}
\end{figure*}

\subsection{Continual learning techniques}\label{sec:CL_technique}
 

Existing CL techniques to handle catastrophic forgetting can be broadly classified into three categories \cite{li2020continual}\cite{maltoni2019continuous}, viz., rehearsal \cite{lopez2017gradient}\cite{chaudhry2018efficient}\cite{rolnick2019experience}\cite{rebuffi2017icarl}, regularization \cite{kirkpatrick2017overcoming}\cite{zenke2017continual}\cite{MALTONI201956}\cite{li2017learning}, and architectural \cite{rusu2016progressive}\cite{yoon2017lifelong}. 
In rehearsal-based techniques, a small memory is used to store previous task data and then replayed along with current data to overcome forgetting. The previous data can be stored and replayed in various forms, including raw samples, deep feature descriptors, and generative models to generate samples or features. In contrast, Regularization-based techniques avoid storing past samples and add a regularization term in the loss function or regularize the learning rate to control drastic deviations in learned weights to minimize forgetting. Further, data-focused and prior-focused regularization have been explored.  Data-focused approaches use knowledge distillation to transfer past learning to the current model, whereas the prior-focused methods define the importance of the network's parameters, which is then used to penalize larger deviation in important parameters~\cite{kirkpatrick2017overcoming}. Architectural techniques mainly aim to keep some network parameters reserved for each episode. Two design choices are explored in this category: fixed network capacity and dynamic network capacity. Lastly, hybrid approaches, i.e., a combination of two or more of the above categories, are also popularly explored in various applications~\cite{chen2019hybrid}. 

\subsection{Continual Learning for Audio Analysis}
Table \ref{tab:comp_literature} summarises various studies that apply CL to audio applications and utilize CL strategies, scenarios, and datasets.
For audio scene classification, researchers have employed CI and TI scenarios to handle the dynamic nature of audio data. The works \cite{l2_wang2019continual}\cite{l3_koh2020incremental}\cite{l8_kwon2021fasticarl}\cite{l9_wang2022learning}\cite{l14_mulimani2023incremental} and \cite{l16_sun2023environmental} focused on CI scenario. In contrast, Karam et al. \cite{l10_karam2022task} targeted the TI scenario for classification.
In multi-label audio classification \cite{l4_wang2021few}\cite{l7_wang2021calls}\cite{l17_mulimani2024class}, researchers have used CI scenario on  AudioSet, ESC50, and FSD-MIX-CLIPS datasets. Authors used CI scenarios for multi-modal classification, which involves audio and video data \cite{l11_mo2023class} \cite{l12_pian2023audio}. They used the VGGSound and AVE-CI datasets to ensure the model could learn from both audio and visual inputs without degrading performance. Kim et al.  \cite{l13_kim2024multi} applied the TI scenario on diverse datasets such as VGGSound, MACS, and ClothoV2 datasets.
 Berg et al. \cite{l6_berg2021continual} used Clotho and AudioCaps datasets to maintain accuracy in generating captions for audio clips. Ma et al. \cite{l5_ma2021continual} focused on the ASVSpoof2019 dataset for fake audio detection, where drift is considered as the different types of spoofing attacks with major differences. Zhou et al. \cite{l18_zhou2024anomaly} used the DI scenario for anomaly sound detection in the DCASE 2020 challenge, addressing the challenge of subsequent machine-type drift and ensuring robust anomaly identification over time.

From the summary provided in Table \ref{tab:comp_literature}, we can observe that most of the works focus on audio scene classification. Further, most ``new class occurrence'' type situation (class-incremental) is considered in these works, whereas the ``drifted data'' situation (domain-incremental), which is a more prevalent and critical issue in outdoor audio scene analysis, is largely ignored. Less exploration is also attributed to the lack of suitable public datasets for the evaluation of a CL framework.

\section{Baselines}\label{sec:baslines}
The CL research community has contributed various CL techniques in the categories mentioned in Section~{\ref{sec:CL_technique}}. Their effectiveness is also evaluated against non-CL approaches that try to mitigate forgetting using retraining or exploiting all the datasets together. However, these approaches do not fit with human-like learning but may perform well at the cost of large storage and computation facilities.  For the current application of audio analysis, we have evaluated the following CL and non-CL baselines.



\subsection{Continual learning baselines}
This paper targets a few major and frequently used CL baselines to conduct a comparative study on the dataset. These baselines are briefly discussed below.    

\subsubsection{Elastic Weight Consolidation} Elastic Weight Consolidation (EWC) serves as a regularization technique designed to mitigate the issue of forgetting in neural networks. It achieves this by selectively constraining, or partially freezing, the model weights associated with essential components for previously learned tasks. By prioritizing the preservation of crucial knowledge, EWC helps maintain performance on earlier tasks while accommodating the learning of new ones. This method provides a balance, preventing excessive interference with existing knowledge when adapting to novel information \cite{MALTONI201956,kirkpatrick2017overcoming}.


\subsubsection{Learning without Forgetting} Learning without Forgetting (LwF) is a regularization method aimed at addressing the challenge of forgetting in neural networks by enforcing stability in the output or predictions. Notably, LwF prioritizes training efficiency, demonstrating faster results compared to the joint training approach. This is a CL approach specifically designed to utilize only new data, assuming the unavailability of past data used for initial network pre-training. This sets LwF apart from other CL methods, which typically focus on leveraging prior knowledge to facilitate the learning of new tasks \cite{MALTONI201956}\cite{li2017learning}.




\subsubsection{Synaptic Intelligence} Synaptic Intelligence (SI) is treated as a variant of EWC. In contrast to EWC, which estimates parameter importance through a diagonal approximation of the Fisher information, synaptic intelligence takes an online approach, dynamically estimating parameter importance based on the ongoing training trajectory. While EWC relies on a fixed approximation, synaptic intelligence adapts continuously during the learning process, providing a more flexible assessment of parameter significance. This distinction underscores the difference in how these methods evaluate and prioritize the importance of parameters in neural network training \cite{MALTONI201956}\cite{zenke2017continual}.


\subsubsection{Gradient Episodic Memory} Gradient Episodic Memory (GEM) serves as a framework for continual learning, aiming to mitigate catastrophic forgetting by utilizing an episodic memory to retain a portion of the current data samples. By adjusting the gradient of each current task, GEM endeavors to minimize negative backward transfer, thereby preventing the loss of information from previously learned tasks. This approach ensures that the knowledge gained from past tasks positively influences the learning of subsequent tasks, contributing to a more effective and stable learning process \cite{lopez2017gradient}\cite{chen2019revisiting}.


\subsubsection{Averaged Gradient Episodic Memory} Averaged Gradient Episodic Memory (A-GEM) tries to alleviate the computational cost caused by GEM. A-GEM can be defined as the more efficient version of GEM. Unlike GEM, which minimizes the loss of each previous task at each training step by adjusting the gradient of each current task, A-GEM minimizes the average episodic memory loss over the previous tasks at each training step \cite{chaudhry2018efficient}.

\subsubsection{Greedy sampler and Dumb learner} In Greedy sampler and Dumb learner (GDumb), the sampler greedily stores the part of data samples as they are encountered while ensuring a balanced representation of classes, and the learner, which is a neural network, trains it dumbly using all the samples available in the memory. Therefore it is called GDumb. Although GDumb is not meant to handle the continual learning scenarios, it gives convincing results in comparison to the CL strategies \cite{prabhu2020gdumb}.

\subsubsection{Replay} The replay strategy can use different types of replay, like, experience replay \cite{rolnick2019experience} where the old stored samples from the memory are used for training with the new samples, generative replay \cite{shin2017continual} where the old data is generated using some generative models and trained with the new encountered data samples etc. 


\subsection{Non-Continual learning baselines}
The effectiveness of CL methods is compared against some non-CL approaches that try to mitigate forgetting and achieve high performance using retraining or exploiting all the episode datasets. These approaches do not fit with human brain learning but may perform well at the cost of large amounts of space and computation facilities. We discuss them below.


\subsubsection{Joint-training} This is a popular approach in the non-CL category, where all the training episodes are jointly trained one time and achieve the highest average performance and hence also regarded as an upper bound. However, this method assumes the availability of all the episodes simultaneously, which is a bottleneck.


\subsubsection{Cumulative learning} 
This non-CL training strategy is another widely adopted approach offering upper-bound performance. Here, the training process is started from scratch each time a new episode is available. It considers the current and all the previous episodes together for retraining. 
However, previously learned knowledge is wasted due to retraining from scratch each time a new episode is encountered, leading to high computation requirements.


\subsubsection{Naive} This non-CL strategy achieves lower-bound performance. Here, a traditional transfer learning mechanism is followed. Whenever a new episode is available, the model adapts itself to this new data without following any mechanism to control forgetting past data.



\begin{figure*}[!ht]
\centering
\includegraphics[scale=0.59]{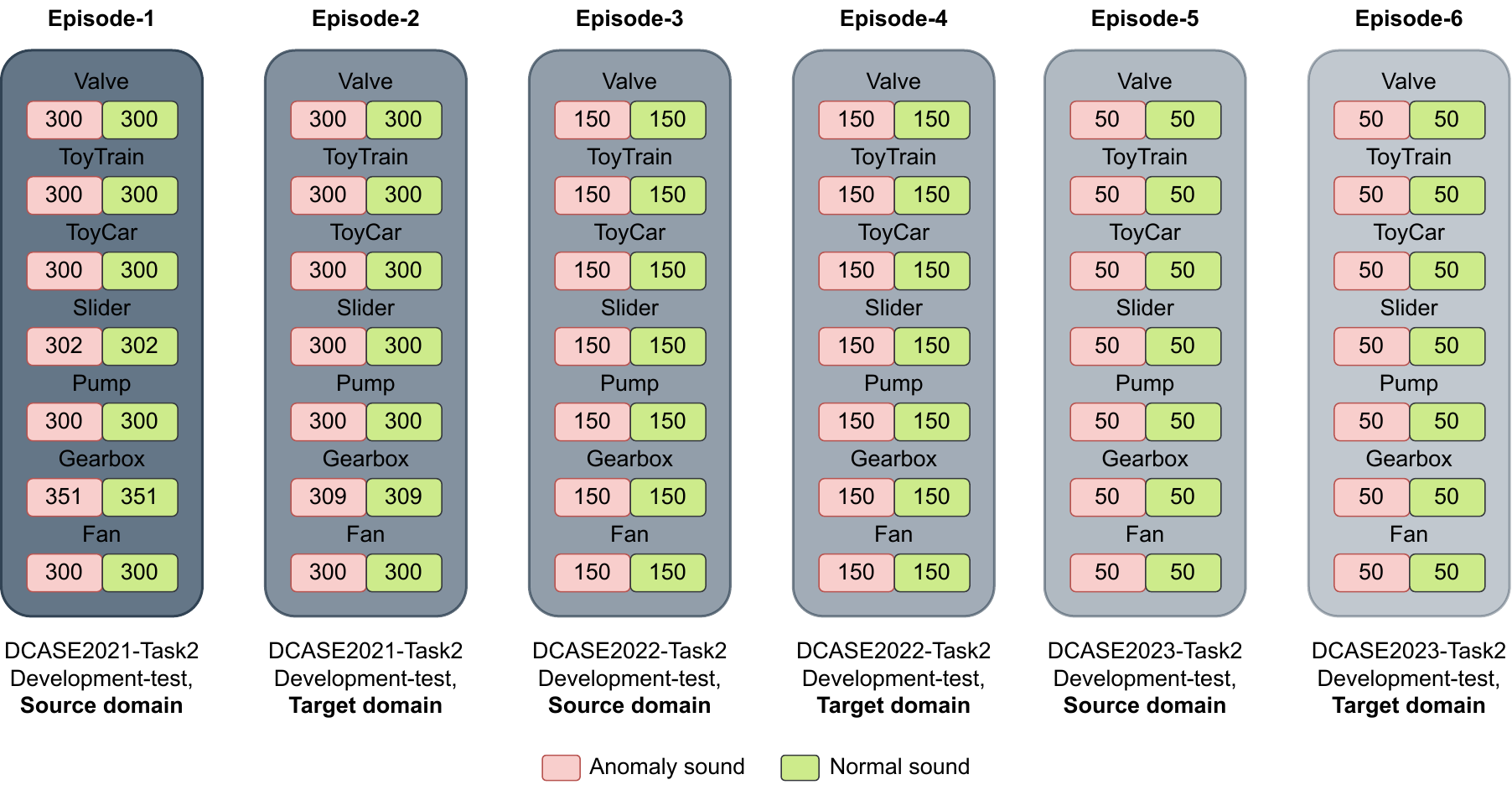}
\caption{Benchmarking dataset for DI scenario. Here the CL model aims to distinguish between normal and abnormal machine sounds in domain shift conditions.}
\label{fig:DISdemo}
\end{figure*}
\begin{figure*}[!htbp]
\centering
\includegraphics[scale=0.57]{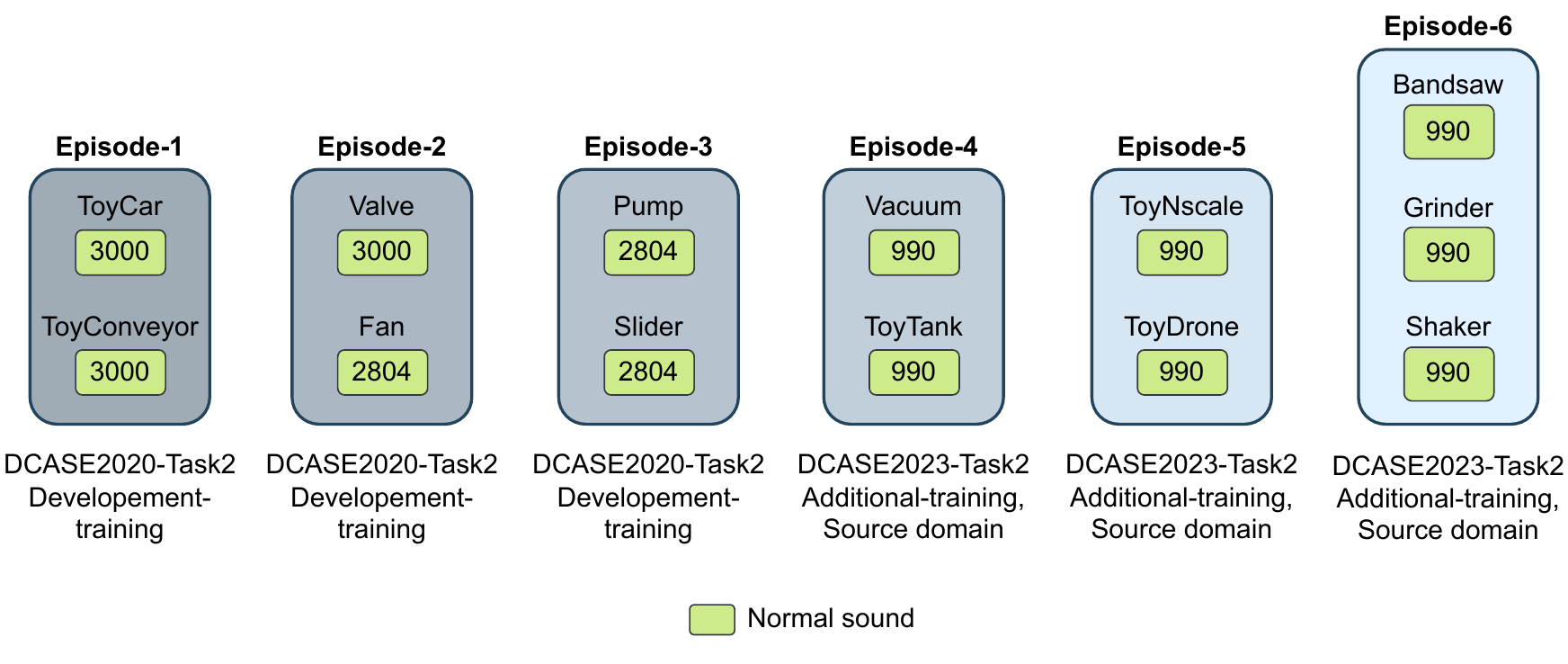}
\caption{Benchmarking dataset for CI scenario. Here the classification model aims to distinguish among machine types in class incremental fashion. Here only normal sound of considered machine types is available.}
\label{fig:CISdemo}
\end{figure*}
\begin{table*}[!htbp]
\centering
\caption{Details of datasets}
\label{tab:datasets_samples}

\begin{tabular}{|c|c|c|c|c|c|}
\hline
{\bf Continual Learning Scenario} &
{\bf Application}&
  {\bf Tasks} &
  {\bf \#Train clips} &
  \begin{tabular}[c]{@{}c@{}}{\bf \#Test clips}\end{tabular} \\ \hline

  \begin{tabular}[c]{@{}c@{}} Domain-incremental \end{tabular} &
  \begin{tabular}[c]{@{}c@{}} Anomalous \\sound detection\end{tabular}&
  \begin{tabular}[c]{@{}c@{}}  T1 (DCASE2021 source), \\T2 (DCASE2021 target),\\ T3 (DCASE2022 source), \\T4 (DCASE2022 target), \\T5 (DCASE2023 source),\\T6 (DCASE2023 target) \end{tabular}&
  \begin{tabular}[c]{@{}c@{}} 3098, \\3036, \\1512,\\1512,\\504,\\504  \end{tabular}& 
  \begin{tabular}[c]{@{}c@{}} 862,\\844,\\420,\\420,\\140,\\140   \end{tabular}\\ \hline

  \begin{tabular}[c]{@{}c@{}} Class-incremental \end{tabular} &
  \begin{tabular}[c]{@{}c@{}} Machine \\sound identification\end{tabular}&
  \begin{tabular}[c]{@{}c@{}}  T1 (ToyCar, ToyConveyor), 
  \\T2 (Valve, Fan), \\T3 (Pump, slider), \\T4 (Vacuum, ToyTank), \\T5 (ToyNscale, ToyDrone), \\T6 (Bandsaw, Grinder, Shaker) \end{tabular}&
  \begin{tabular}[c]{@{}c@{}} 4320,\\ 4178, \\4037, \\1425, \\1425, \\2138  \end{tabular}& 
  \begin{tabular}[c]{@{}c@{}} 1200,\\ 2361,\\ 3483, \\3879, \\4275, \\ 4869  \end{tabular}\\ \hline 
   
\end{tabular}
\end{table*}

\section{Datasets}\label{sec:dataset}
To systematically evaluate the CL techniques for monitoring, we require a sequence of datasets to simulate CL scenarios. To the best of our knowledge, no such benchmark datasets have been proposed for audio analysis. Further, there is no large public audio dataset for monitoring; however, there is a well-defined and gigantic repository for abnormality detection in machine sound, namely DCASE. This is very close to the audio-based monitoring application and, hence, suitable for benchmarking CL techniques on audio analysis. Moreover, audio clips comprised various machine sounds in the normal and abnormal categories with and without shifted data distribution due to different machine loads and environmental factors. We thus curate CL benchmark datasets representing different CL scenarios from publically available audio datasets from the DCASE community.

DCASE challenge offers datasets for various audio analysis-related applications every year. Among various tasks, the task-2 of DCASE is for audio anomaly detection and has been held continuously since the year 2020. We have utilized data from the following DCASE challenge tasks from years 2020 to 2023: (a) DCASE2020 task-2: ``Unsupervised Detection of Anomalous Sounds for Machine Condition Monitoring'', (b) DCASE2021 task-2: ``Unsupervised Anomalous Sound Detection for Machine Condition Monitoring under Domain Shifted Conditions", (c) DCASE2022 task-2: ``Unsupervised Anomalous Sound Detection for Machine Condition Monitoring Applying Domain Generalization Techniques", and (d) DCASE2023 task-2: ``First-Shot Unsupervised Anomalous Sound Detection for Machine Condition Monitoring". Each of the tasks offers audio clips from 6-7 distinct machine sounds in normal and abnormal classes. The dataset is provided in three segments: development, evaluation, and additional training. We utilized the development segment from each dataset to curate training and testing episodes. Since class-incremental and domain-incremental are the most frequently observed scenarios, therefore we curate these two CL scenarios as described below.

\subsection{Dataset for DI Scenario}
Here, we consider development datasets provided for task-2 of the DCASE2021, DCASE2022, and DCASE2023 challenges as they have the same machine type but have domain shift conditions due to weather and variable machine load. The development dataset has a source domain and then a target domain with a shifted data distribution, which would serve as the domain-incremental condition for our case. Further, the test part of the development segment only offers balanced normal and abnormal classes, and hence, we consider only the test part, not the training part of the development segment. These datasets have normal and abnormal audio clips for each of the 7 machine types, including Fan, Gearbox, Pump, Slider, ToyCar, ToyTrain, and Valve. The end goal here is abnormal sound detection in a sequence of data that may contain a distribution shift over time. We curate six episodes by utilizing source and target domain data from DCASE2021, DCASE2022, and DCASE2023. The first and second episodes represent the source and the target domains of DCASE2021, having 4306 (2153 normal and 2153 abnormal) and 4218 (2109 normal and 2109 abnormal) samples, respectively. Then, the third and fourth episodes are curated from the source and target domain of DCASE2022, offering 2100 (1050 normal and 1050 abnormal) and 2100 (1050 normal and 1050 abnormal) samples, respectively. Lastly, the fifth and sixth episodes are curated from the source and target domain of the DCASE2023 challenge, which has 700 (350 normal and 350 abnormal) and 700 (350 normal and 350 abnormal) samples, respectively. A pictorial representation is provided in Fig.~\ref{fig:DISdemo} for better visualization of the sequence of episodes under the data-incremental scenario.
\subsection{Dataset for CI Scenario}

Contrary to the domain-incremental dataset discussed in the earlier section, here, the end goal is machine-type classification. In the class-incremental learning scenario, there needs to be the inclusion of classification classes over time. Hence, we opt to classify machine types as the DCASE datasets offer a large number of machine types, which can then be sequentially learned by the CL model. To focus only on machine-type classification, either normal or abnormal data should be considered. Since the normal class data is available in large amounts, hence we consider the normal class samples of different machine sounds here.
Specifically, we curate six episodes by utilizing the training part (it has only normal samples) of the development segment in task-2 of DCASE2020 and the source domain of the additional training segment in task-2 of DCASE2023 challenges. The choice of these datasets is such that they have disjoint machine classes.
Specifically, we consider ToyCar and ToyConveyer machine types from DCASE2020 in the first episode, Valve, and Fan machine types in the second episode, and then Pump and Slider machine types in the third episode.
Then, from the DCASE2023, two machine types, viz., Vacuum and ToyTank, are considered in the fourth episode, ToyNscale and ToyDrone are considered in the fifth episode, and finally, the last episode has the remaining classes from DCASE2023, including Bandsaw, Grinder, and Shaker. Inside every episode, the number of samples for each class is considered in balance amount, i.e., the machine types have 3000, 2804, 990, 990, and 990 samples in the first, third, fourth, fifth, and sixth episodes, respectively. Except for the second episode in which the two machine types, namely Valve and Fan, have 3000 and 2804 samples, respectively. A visual representation of the class-incremental dataset is depicted in Fig.~\ref{fig:CISdemo}. The detailed description of datasets of both scenarios is summarized in Table \ref{tab:datasets_samples}.

\section{Experiments}\label{sec:experiments}
\subsection{Experimental setup}
In all the approaches, we keep the base model as a pre-trained ResNet50~\cite{he2016deep} model initialized with ImageNet weights. We used Adam as an optimizer. In order to select the best hyperparameters for the classification model, multiple experiments were performed with a cumulative approach as it gives the upper bound. Specifically, we search best values for epochs in \{10, 15,...50\}, batch sizes in \{4,8,...32\}, and learning rates in \{$1e-04$, ... $1e-05$\}. Best performance is achieved with a batch size of 8, a learning rate of $1e-04$, and epochs of 30 for the CI scenario. For the DI scenario, the best performance is achieved with a batch size of 8, a learning rate of $1e-03$, and epochs of 50. To facilitate a common evaluation scheme, the selected hyperparameters are used in all the approaches. Further, the hyperparameters for CL approaches, such as $\lambda$, memory size, etc., were also experimentally selected from a range of values. Specifically, for DI experiment, we set $\lambda=0.5$ in EWC, \{$\alpha=2$, $T=2$\} in LwF, and $\lambda=0.8$ in SI. 
In case of CI experiment, $\lambda=2$ was set in EWC, \{$\alpha=2$, $T=2$\} kept for LwF, and $\lambda=2$ for SI.
Further, for both DI and CI, the memory sizes were kept as 2000 samples for the GDumb and Replay approaches. 
Audio clips were trimmed to 10 seconds, and then spectrograms were computed for each sample.

\begin{figure}[!htbp]
\centering
\includegraphics[scale=0.7]{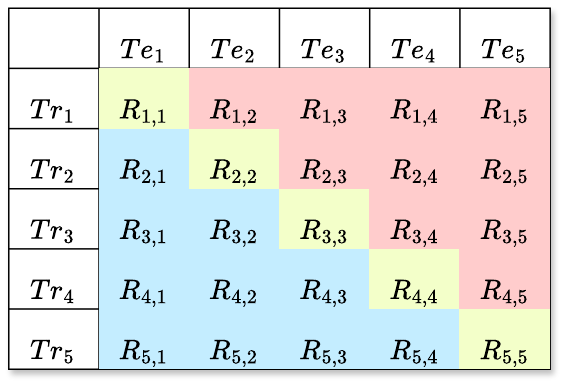}
\caption{The train-test matrix for T=5. }
\label{fig:trainTestDemo}
\end{figure}
\subsection{Evaluation metrics}
Here, we describe the evaluation procedure and metrics in CL. Each of the training sessions is a classification problem; therefore, we compute accuracy on the testing data of each task. After completing the training session of $t^{th}$ task using training data $Tr_t$, we compute accuracy values on testing datasets $Te_1, Te_2, \dots$ yielding $T$ accuracy values. Thus, after completing all the training sessions on a stream of $T$ tasks, we get a train-test matrix $R^{T \times T}$, where $R_{t,j}$ represents accuracy on $j^{th}$ task after completing the training on $t^{th}$ task. An example of a train-test matrix with 5 tasks is shown in Fig.~\ref{fig:trainTestDemo}.

Apart from getting average accuracy across tasks, other important evaluation measures in CL include forgetting and forward transfer, which can be computed from this $R$ matrix. Below, we describe metrics for forward transfer, forgetting, and model accuracy.



\textbf{Backward Transfer (BWT):} BWT, a forgetting measure, quantifies the amount of influence of the current task on the previously learned task~\cite{lopez2017gradient}. It shows the improvement in the performance of a task after learning new tasks. Thus, a positive value of BWT signifies that learning the current task $t$ has caused improvement in the performance of an old task $k<t$. On the other hand, a negative value of BWT indicates that learning the current task $t$ has caused degradation in the performance of an old task. Thus, an algorithm with greater values of BWT is superior to the one with small BWT. Blue and yellow cell values in Fig.~\ref{fig:trainTestDemo} are utilized to compute BWT using the equation \cite{diaz2018don} below.

\begin{equation}
   BWT= \frac{2}{T(T-1)} \sum_{i=2}^{T} \sum_{j=1}^{i-1} (R_{i,j}-R_{j,j})
\end{equation}


\textbf{Forward Transfer (FWT):} FWT signifies the amount of influence of the current task on upcoming tasks. It shows the zero-shot learning capabilities of the CL model. The higher the value of FWT, the better the model. Red cell values in Fig.~\ref{fig:trainTestDemo} are utilized to compute FWT~\cite{diaz2018don} measure with the following equation.

\begin{equation}
FWT=\frac{2}{T(T-1)}  \sum_{t<j}^T {R_{t,j}} 
\end{equation}


\textbf{Accuracy:} Given the train-test matrix $R^{T \times T}$, we compute average accuracy, $ACC$~\cite{lopez2017gradient} across tasks after completing the final task $T$ as below:
\begin{equation}
   ACC=  \frac{1}{T}\sum_{j=1}^{T} R_{T,j}   
\end{equation}

D{\'\i}az-Rodr{\'\i}guez et. al~\cite{diaz2018don} defines a more rigorous measure of average accuracy by computing $ACC$ after completing each task, termed as $A$. It considers model performances at every training session $t$ and thus better represents the dynamic aspects of CL. 

\begin{equation}
   A= \frac{2}{T(T+1)} \sum_{i\geq j}^{T} R_{i,j}
\end{equation}

\begin{figure}[!htbp]
\centering
\includegraphics[scale=0.48]{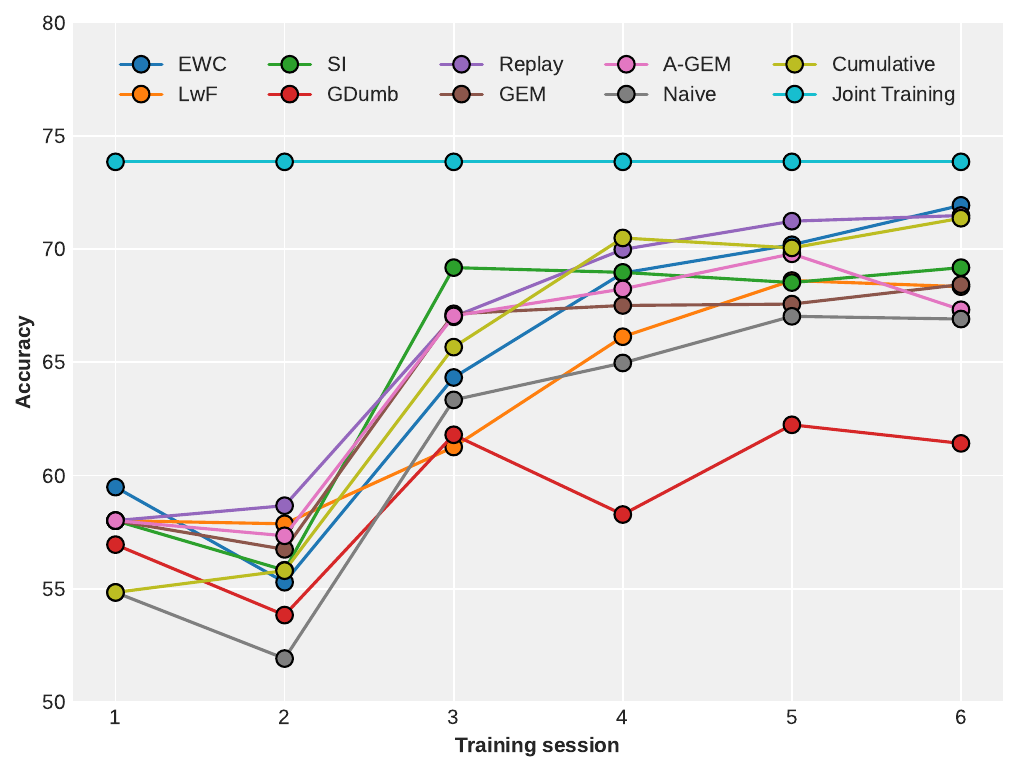}
\caption{Comparison of different approaches in terms of average accuracy over all episodes after each training session in DI scenario.} 
\label{fig:DIS_graph}
\end{figure}

\begin{table}[htbp!]
\caption{Performance comparison for \textbf{DI scenario}. First / second best performance in CL categories indicated in \textcolor{red}{red} / \textcolor{blue}{blue}, respectively. {\bf Bold:} Upper bound.}
\centering

\label{tab:DIL}

\begin{tabular}{|c|c|c|c|c|c|}
\hline
&\multicolumn{1}{|c|}{\textbf{Approach}} & \multicolumn{1}{c|}{\textbf{\begin{tabular}[c]{@{}c@{}}BWT\\ ~\cite{diaz2018don}\end{tabular}}} & \multicolumn{1}{c|}{\textbf{\begin{tabular}[c]{@{}c@{}}FWT\\ ~\cite{diaz2018don}\end{tabular}}} & \multicolumn{1}{c|}{\textbf{A\cite{diaz2018don}}} & \multicolumn{1}{c|}{\textbf{\begin{tabular}[c]{@{}c@{}}ACC\\ ~\cite{lopez2017gradient}\end{tabular}}} \\ \hline



\multirow{7}{*}{\rotatebox{90}{CL}} &\begin{tabular}[c]{@{}c@{}}EWC ($\lambda=2$)\end{tabular} & -5.22 & 65.09 &\textcolor{blue}{67.67} & \textcolor{red}{71.93} \\ \cline{2-6}
&\begin{tabular}[c]{@{}c@{}}LwF (${\alpha=1, T=1}$)\end{tabular} & \textcolor{red}{-1.23} & \textcolor{blue}{65.85} & 66.87 & 68.34  \\\cline{2-6}
&\begin{tabular}[c]{@{}c@{}}SI ($\lambda=0.5$)\end{tabular} & -9.44 & 63.92 & 67.31 & 69.18  \\ \cline{2-6}
&\begin{tabular}[c]{@{}c@{}}GDumb (mem=2000)\end{tabular} & -2.25 & 61.55 & 61.60 & 61.42  \\ \cline{2-6}
&\begin{tabular}[c]{@{}c@{}}Replay (mem=2000)\end{tabular} &\textcolor{blue}{ -2.22 }& \textcolor{red}{68.98} & \textcolor{red}{70.12}  & \textcolor{blue}{71.48 } \\ \cline{2-6}
&\begin{tabular}[c]{@{}c@{}}GEM (mem=200*6)\end{tabular} & -7.40 & 64.03  & 66.70  & 68.43   \\ \cline{2-6}
&\begin{tabular}[c]{@{}c@{}}A-GEM (mem=200*6)\end{tabular} & -6.83 & 64.91 & 67.28 & 67.32   \\ \hline

\multirow{3}{*}{\rotatebox{90}{Non-CL}} &Naive  & -9.44  & 63.91  & 67.31  & 69.18  \\ 
&Cumulative & \textbf{-0.17} &\textbf{ 71.23} &\textbf{ 72.15} & 75.40  \\ 
&Joint & --   & -- & --  & \textbf{76.81}  \\ \hline
\end{tabular}

\end{table}

\subsection{Results}
\subsubsection{Performance analysis for DI scenario}
For the DI scenario, we compute average accuracy over all tasks after completing a training session and report the results via Fig~\ref{fig:DIS_graph}. We can see that the joint training approach gives the upper bound of performance. CL approaches perform similarly. Further, the GDumb strategy and naive training strategy give the lowest performance compared to others. Naive training strategy does not consider any measure to handle forgetting of past tasks and hence lower average performance. On the other hand, GDumb randomly considers a subset of data from all encountered tasks for learning from scratch and hence performs poorly due to sample selection bias. 

A detailed comparison of the best performance achieved by various approaches is provided in Table~\ref{tab:DIL}. Specifically, for each approach, we report forgetting measures by BWT, forward transfer by FWT, incremental learning capabilities by $A$, and final average accuracy by $ACC$. Here, we can also see that joint and cumulative approaches in the non-CL category tend to outperform in all performance metrics as compared to CL approaches. However, the assumption of availability of all the tasks together and retraining from scratch whenever a new task arrives limits their applicability in real life. While comparing the CL approaches, we can see that LwF better handles forgetting achieving the best BWT of -1.23. However, other metrics (FWT, $A$, and $ACC$) are lower compared to that achieved by Replay. This is due to the fact that the knowledge distillation step in LwF limits the plasticity to learn new tasks. It usually performs well in situations involving less drift/ new data over the episodes (also evident from compromised results in CI scenario). Remarkably, replay outperforms other CL approaches, including LwF, by achieving (FWT, $A$, and $ACC$) as $\{68.98, 70.12, 71.48\}$. Further, we can see that GDumb, with the same memory size as Replay, performs poorly. GEM and A-GEM approaches are highly compute-intensive compared to Replay and achieve lower performance than replay. However, these CL approaches do not show a contrasting superiority over another DI scenario. Therefore, CL approaches may be picked based on the application requirements; for example, a system demanding a solution with low memory will favor memory-free approaches such as EWC, LwF, or SI. Otherwise, way can favor the Replay approach as it offers the best overall performance.

\begin{figure}[!htbp]
\centering
\includegraphics[scale=0.45]{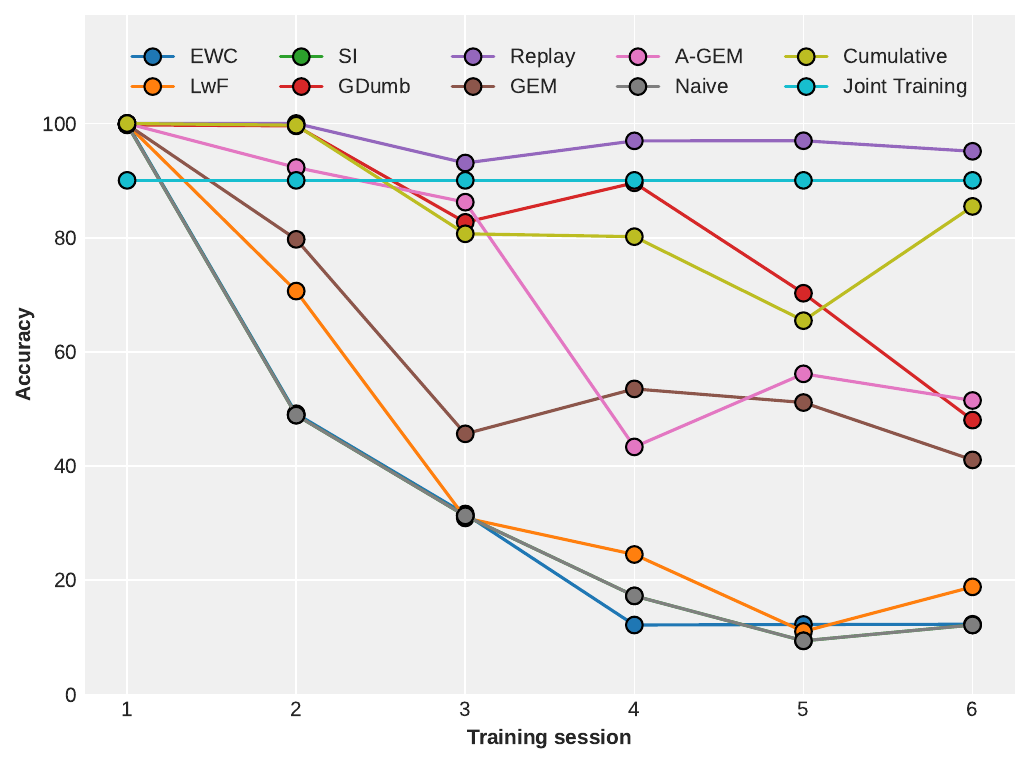}
\caption{Comparison of different approaches in terms of average accuracy after each training session in CI scenario.}
\label{fig:CIS_graph}
\end{figure}

\begin{table}[!htbp]
\caption{Performance Comparison for \textbf{CI scenario}. First / second best performance in CL categories indicated in \textcolor{red}{red} / \textcolor{blue}{blue} , respectively. {\bf Bold:} Upper bound.}
 \label{tab:CIL}
 \centering
\begin{tabular}{|c|c|c|c|c|c|}
\hline

&\multicolumn{1}{|c|}{\textbf{Approach}} & \multicolumn{1}{c|}{\textbf{\begin{tabular}[c]{@{}c@{}}BWT\\ ~\cite{diaz2018don}\end{tabular}}} & \multicolumn{1}{c|}{\textbf{\begin{tabular}[c]{@{}c@{}}FWT\\ ~\cite{diaz2018don}\end{tabular}}} & \multicolumn{1}{c|}{\textbf{A\cite{diaz2018don}}} & \multicolumn{1}{c|}{\textbf{\begin{tabular}[c]{@{}c@{}}ACC\\ ~\cite{lopez2017gradient}\end{tabular}}} \\ \hline

\multirow{7}{*}{\rotatebox{90}{CL}} &\begin{tabular}[c]{@{}c@{}}EWC ($\lambda=2$)\end{tabular} & -54.60 & 0.61 & 10.79 & 2.07 \\  \cline{2-6}
&\begin{tabular}[c]{@{}c@{}}LwF (${\alpha=2, T=2}$)\end{tabular} & -45.72 & 12.35 & 20.28 & 2.96 \\ \cline{2-6}
&\begin{tabular}[c]{@{}c@{}}SI ($\lambda=2$)\end{tabular} & -54.77 & 0 & 10.13 & 2.03 \\ \cline{2-6}
&\begin{tabular}[c]{@{}c@{}}GDumb (mem=2000)\end{tabular} & \textcolor{blue}{-17.58} & \textcolor{blue}{73.39} & \textcolor{blue}{77.18} & \textcolor{blue}{53.15} \\ \cline{2-6}
&\begin{tabular}[c]{@{}c@{}}Replay (mem=2000)\end{tabular} & \textcolor{red}{\textbf{-1.04}} & \textcolor{red}{\textbf{96.97}} & \textcolor{red}{\textbf{96.98}} &\textcolor{red}{ \textbf{95.74} }\\\cline{2-6}
&\begin{tabular}[c]{@{}c@{}}GEM (mem=200*6)\end{tabular} & -39.87 & 34.34 & 42.19 & 33.01 \\\cline{2-6}
&\begin{tabular}[c]{@{}c@{}}A-GEM (mem=200*6)\end{tabular} & -22.61 & 62.11 & 64.82 & 50.93 \\ \hline

\multirow{3}{*}{\rotatebox{90}{Non-CL}} &Naive & -54.96 & 0.55 & 10.81 & 2.03 \\ 
&Cumulative & -4.93 & 86.18 & 85.92 & 88.22 \\ 
&Joint & -- & -- & -- & 90.04 \\ \hline
\end{tabular}
\end{table}

\subsubsection{Performance analysis for CI scenario}
Figure \ref{fig:CIS_graph} shows the performance of the different approaches in terms of average accuracy on only seen classes after learning each task in the CIS setting. It is expected that over the session, performance would drop as more and more classes are added, making learning difficult. However, Replay is able to maintain the performance to a steady level over the training session. We can see that it outperforms all other CL approaches, even surpassing non-CL strategies that are supposed to give an upper-bound performance, such as cumulative and joint. 

We report a detailed comparison of all approaches in terms of BWT, FWT, $A$, and $ACC$ in Table \ref{tab:CIL}. 
From the table, we can observe that regularization-based approaches (EWC, LwF, SI) perform similarly to the naive approach. We can observe that the Replay approach achieves the best values of (BWT, FWT, $A$, $ACC$) as $\{-1.04, 96.97, 96.98, 95.74\}$. It outperforms with an incremental learning accuracy (A) of 96.98\% compared to other rehearsal approaches like GDumb, GEM, and A-GEM, which achieve only 77.18\%, 42.19\%, and 64.82\%, respectively. We can see that GDumb, which is not a carefully designed rehearsal-based approach, achieves the second-best results and shows superior performance to complex rehearsal-based approaches like GEM and A-GEM. 

From tables \ref{tab:DIL} and \ref{tab:CIL}, we conclude that the rehearsal-based strategies perform well in both DI and CI scenarios, while regularization approaches show inadequate performance in the case of CI scenarios. This is attributed to the fact that the CI scenario is the most difficult scenario as the new task consists of novel classes, and hence, it is very hard to learn new tasks along with remembering the past just with the regularization step in the loss function. Some samples from past classes are important to rejuvenate the information from past classes in the current model.




\begin{table}[!ht]
\caption{Performance Comparison with different hyperparameters for \textbf{DI scenario}.}

\label{tab:DIL_2}
\centering
\begin{tabular}{|l|l|l|l|l|}
\hline
\multicolumn{1}{|c|}{\textbf{Approach}} & \multicolumn{1}{c|}{\textbf{\begin{tabular}[c]{@{}c@{}}BWT\\ ~\cite{diaz2018don}\end{tabular}}} & \multicolumn{1}{c|}{\textbf{\begin{tabular}[c]{@{}c@{}}FWT\\ ~\cite{diaz2018don}\end{tabular}}} & \multicolumn{1}{c|}{\textbf{A\cite{diaz2018don}}} & \multicolumn{1}{c|}{\textbf{\begin{tabular}[c]{@{}c@{}}ACC\\ ~\cite{lopez2017gradient}\end{tabular}}} \\ \hline

 \rowcolor{gray!10}EWC ($\lambda=0.5$) & -7.85 & 64.51 & 67.03 & 66.99 \\ 
 \rowcolor{gray!10}EWC ($\lambda=1$) & -6.65 & 63.86 & 66.79 & 68.46   \\ 
 \rowcolor{gray!10}EWC ($\lambda=2$) &\textbf{-5.22} & \textbf{65.09} & \textbf{67.67} & \textbf{71.93}  \\ \hline

LwF (${\alpha=1, T=1}$) & \textbf{-1.23} & \textbf{65.85} & \textbf{66.87} & \textbf{68.34}   \\ 
LwF (${\alpha=2, T=1}$) & -1.59 & 62.65 & 62.65 & 63.10   \\ 
LwF (${\alpha=3, T=1}$) & -2.87 & 60.03 & 60.97 & 58.72  \\ \hline

 \rowcolor{gray!10}SI ($\lambda=0.5$) & -9.44 & \textbf{63.92} & \textbf{67.31} & \textbf{69.18}   \\ 
 \rowcolor{gray!10}SI ($\lambda=1$) & \textbf{-7.86} & 63.44 & 66.50 & 68.20   \\ 
 \rowcolor{gray!10}SI ($\lambda=2$) & \textbf{-7.86} & 63.44 & 66.50 & 68.20  \\ \hline

GDumb (mem=2000) & -2.25 & \textbf{61.55} & \textbf{61.60} & \textbf{61.42} \\ 
GDumb (mem=1500) & \textbf{-0.64} & 60.21 & 60.14 & 58.69 \\ 
GDumb (mem=1000) & -2.36 & 57.44 & 57.82 & 56.62 \\ \hline

 \rowcolor{gray!10}Replay (mem=2000)  & \textbf{-2.22} & \textbf{68.98} & \textbf{70.12} & \textbf{71.48} \\
 \rowcolor{gray!10}Replay (mem=1500)  & -2.40 & 68.71 & 69.74 & 70.80  \\
 \rowcolor{gray!10}Replay (mem=1000)  & -3.65 & 67.69 & 69.36 & 70.31  \\
 \rowcolor{gray!10}Replay (mem=500)  & -3.88 & 66.69 & 68.49 & 70.61 \\
 \rowcolor{gray!10}Replay (mem=100)  & -7.17 & 64.37 & 67.02 & 68.61  \\ 
 \rowcolor{gray!10}Replay (mem=50)  & -6.67 & 64.06 & 66.77 & 68.00  \\ \hline

GEM (mem=10*6) & -6.79 & 63.97 & 66.68 & 68.46 \\
GEM (mem=50*6) & \textbf{-5.12} & 63.93 & 66.25 & \textbf{69.88} \\
GEM (mem=200*6) & -7.40 & \textbf{64.03} & \textbf{66.70} & 68.43 \\ \hline

 \rowcolor{gray!10}
A-GEM (mem=10*6) & \textbf{-3.72} & 64.68 & 66.70 & 68.99 \\
 \rowcolor{gray!10}
 A-GEM (mem=50*6) & -2.66 & 65.13 & 67.07 & \textbf{69.12} \\
 \rowcolor{gray!10}
 A-GEM (mem=200*6) & -6.83 & \textbf{64.91} & \textbf{67.28} & 67.32  \\ \hline


\end{tabular}
\end{table}
\begin{table}[!ht]
\caption{Comparison of approaches on different hyperparameters for \textbf{CI scenario}.}

\label{tab:CIL_2}
\centering
\begin{tabular}{|l|l|l|l|l|}
\hline
\multicolumn{1}{|c|}{\textbf{Approach}} & \multicolumn{1}{c|}{\textbf{\begin{tabular}[c]{@{}c@{}}BWT\\ ~\cite{diaz2018don}\end{tabular}}} & \multicolumn{1}{c|}{\textbf{\begin{tabular}[c]{@{}c@{}}FWT\\ ~\cite{diaz2018don}\end{tabular}}} & \multicolumn{1}{c|}{\textbf{A\cite{diaz2018don}}} & \multicolumn{1}{c|}{\textbf{\begin{tabular}[c]{@{}c@{}}ACC\\ ~\cite{lopez2017gradient}\end{tabular}}} \\ \hline

 \rowcolor{gray!10}EWC ($\lambda=1$) & -54.74 & 0.27 & 10.44 & 2.03 \\ 
 \rowcolor{gray!10}EWC ($\lambda=2$) & \textbf{-54.60} & \textbf{0.61} & \textbf{10.79} & \textbf{2.07} \\ 
 \rowcolor{gray!10}EWC ($\lambda=3$) & -54.61 & 0.06 & 10.17 & 2.03 \\ \hline

LwF (${\alpha=1, T=1}$) & -48.68 & \textbf{13.62} & \textbf{21.91} & \textbf{14.31} \\ 
LwF (${\alpha=2, T=2}$) & \textbf{-45.72} & 12.35 & 20.28 & 2.96 \\ 
LwF (${\alpha=3, T=3}$) & -50.42 & 9.63 & 18.39 & 11.39 \\ \hline

 \rowcolor{gray!10}SI ($\lambda=1=3$) & -54.96 & \textbf{0.55} & \textbf{10.81} & 2.03 \\ 
 \rowcolor{gray!10}SI ($\lambda=2$) & \textbf{-54.77} & 0 & 10.13 & 2.03 \\ \hline

GDumb (mem=2000) & -17.58 & 73.39 & \textbf{77.18} & \textbf{53.15} \\ 
GDumb (mem=1500) & -18.06 & 71.08 & 72.92 & 52.35 \\ 
GDumb (mem=1000) & \textbf{-12.95} & \textbf{73.59} & 74.14 & 52.13 \\ \hline

 \rowcolor{gray!10}Replay (mem=2000) & \textbf{-1.04} & \textbf{96.97} & \textbf{96.98} & \textbf{95.74} \\ 
 \rowcolor{gray!10}Replay (mem=1500) & -1.92 & 94.75 & 94.86 & 93.58 \\ 
 \rowcolor{gray!10}Replay (mem=1000) & -1.88 & 94.50 & 94.58 & 93.53 \\ 
 \rowcolor{gray!10}Replay (mem=500) & -2.87 & 94.01 & 94.30 & 92.11 \\ \hline

GEM (mem=200*6) & \textbf{-39.87} & \textbf{34.34} & \textbf{42.19} & 33.01 \\ 
GEM (mem=300*6) & -44.82 & 33.28 & 41.87 & \textbf{39.00} \\ 
GEM (mem=400*6) & -53.92 & 21.39 & 31.76 & 11.75 \\ \hline

 \rowcolor{gray!10}A-GEM (mem=200*6) & \textbf{-22.61} & \textbf{62.11} & \textbf{64.82} & 50.93 \\ 
 \rowcolor{gray!10}A-GEM (mem=300*6) & -26.50 & 59.73 & 63.34 & 39.73 \\
 \rowcolor{gray!10}A-GEM (mem=400*6) & -23.30 & 60.43 & 63.29 & \textbf{54.45} \\ \hline

\end{tabular}
\end{table}

\section{Discussion}\label{sec:dicsussion}

From Table \ref{tab:DIL_2}, we observe that different CL strategies perform equally well in DI scenario. Among the regularization approaches (EWC, LwF, and SI), EWC achieves better performance (BWT, FWT, $A$, $ACC$) as the $\lambda$ value increases, while SI performs better at lower $\lambda$ values. LwF shows an increase in performance (BWT, FWT, $A$, $ACC$) as the $\alpha$ value decreases. EWC outperforms the other regularization approaches in terms of overall accuracy ($A$=67.67\%) as well as average accuracy ($ACC$=71.93\%), while LwF excels in reducing forgetting (BWT=-1.23) and improving future learning (FWT=65.85).

For rehearsal-based approaches (GDumb, Replay, GEM, and A-GEM), we can see that we can improve performance at the cost of increasing memory size. GEM and A-GEM achieve similar performance to regularization approaches (EWC, SI, and LwF) but better than GDumb. Since the compute requirements in GEM and A-GEM are very high, so regularization approaches can be favoured over these. GDumb with lower memory size (mem=1000) shows very poor performance and improves if we increase memory size to 2000. It achieves $A$ as 57\% and 61.60\% with memory sizes of 1000 and 2000, respectively. In contrast, Replay achieves better results than GDumb even with low memory. It records $A$ as 69.36\% and 70.12\% for memory sizes of 1000 and 2000, respectively.

In the CI scenario, as shown in Table \ref{tab:CIL_2}, we observe that regularization-based approaches (EWC, SI, and LwF) fail to mitigate forgetting, for a wide range of hyperparameters as for EWC and SI ($\lambda$ in \{1, 2, 3\}) and for LwF ($\{\alpha=1, T=1\}$, $\{\alpha=2, T=2\}$, $\{\alpha=3, T=3\}$). 
Conversely, rehearsal techniques perform relatively well.
The performance should increase with the increase in memory size, but it is sometimes violated in CI scenarios by GDumb, GEM, and A-GEM. CI scenario is very hard compared to DI, and thus, a carefully designed strategy succeeds in handling the forgetting. We can observe that GDumb performs relatively better compared to GEM and A-GEM. Still, Replay outperforms these approaches by a large margin in terms of all four metrics. On a wide range of memory values, the Replay approach consistently outperforms all other techniques used in the study. The best performance in terms of final accuracy $ACC$ by Replay is 95.74\% whereas that with GDumb, GEM, and A-GEM is 53.15\%, 39.00\%, and 54.45\%, respectively. 
Further, performance by the Replay approach shows a very small decrement even when the memory size is reduced from 2000 samples to 500 samples. 
A-GEM performs better than GEM in terms of both accuracy and training time when varying the number of samples per task (200, 300, 400). Additionally, the training time for Replay with a memory size of 2000 is significantly less than that of GEM and A-GEM with smaller memory sizes of 1200 (200*6) samples.

Overall, Replay emerges as the most effective technique for CI scenarios, demonstrating consistent performance and efficiency across different memory sizes.

\section{Conclusion}\label{sec:conclusion}
We investigate CL approaches on DI and CI scenarios with 6 sequential tasks curated from a large range of DCASE datasets for machine sound monitoring. With the extensive comparative analysis performed, we observe that the performance of rehearsal-based strategies outperforms regularization-based CL approaches in the CI scenario, whereas both categories of approaches perform seemingly well in the DI scenario. We achieve overall accuracy (A) with the Replay approach as 70.12\% for the DI scenario and 96.98\% for the CI scenario. Overall, this comparative analysis emphasizes the importance of selecting appropriate CL strategies adapted to the specific requirements and complexities of audio data, ensuring optimal performance and robustness in real-world applications. 
 
There are different unexplored real-world challenges associated with the use of CL in audio-based applications like machine monitoring or farm surveillance. In the future, we aim to investigate CL in farm monitoring, which is prone to encounter more challenging and diverse domain shifts due to changes in weather, season, and dynamic background environment. Further, in real-life monitoring, investigating CL with blurred domain boundaries is yet to be explored.




\section*{Acknowledgment}
This work is supported by the grant received from DST, Govt. of India for the Technology Innovation Hub at the IIT Ropar in the framework of the National Mission on Interdisciplinary Cyber-Physical Systems.

\bibliographystyle{ieeetr} 
\bibliography{8_bibtex}

\end{document}